\documentclass[11pt]{article}
\usepackage[margin=1.25in]{geometry}
\usepackage{amsmath}
\usepackage{amssymb}
\usepackage{amsthm}
\usepackage{graphicx} %
\usepackage{fullpage}
\usepackage{hyperref}
\usepackage[capitalise]{cleveref}
\usepackage{xcolor}
\usepackage{multicol}
\usepackage{url}
\usepackage[utf8]{inputenc}
\usepackage{enumitem}
\usepackage[bottom]{footmisc}

\newcommand{\card}[1]{\lvert #1\rvert}

\newcommand{\CSP}{\operatorname{CSP}}

\newcommand{\Span}{\operatorname{span}}

\newcommand{\F}{\mathbb{F}}

\newcommand{\OR}{\operatorname{OR}}

\newcommand{\NRD}{\operatorname{NRD}}

\newcommand{\BCK}{\operatorname{BCK}}

\newcommand{\BoolBCK}{\operatorname{BoolBCK}}
\newcommand{\BoolBCKp}{\operatorname{BoolBCK}^{+}}
\newcommand{\EQ}{\operatorname{EQ}}

\newcommand{\bF}{\mathbb{F}}
\newcommand{\bR}{\mathbb{R}}
\newcommand{\R}{\mathbb{R}}

\newcommand{\email}[1]{\href{mailto:#1}{#1}}

\newtheorem{theorem}{Theorem}
\numberwithin{theorem}{section}
\newtheorem{lemma}[theorem]{Lemma}

\newtheorem{proposition}[theorem]{Proposition}
\newtheorem{corollary}[theorem]{Corollary}

\theoremstyle{definition}
\newtheorem{definition}[theorem]{Definition}
\newtheorem{remark}[theorem]{Remark}

\AddToHook{env/lemma/begin}{\crefalias{theorem}{lemma}}
\AddToHook{env/claim/begin}{\crefalias{theorem}{claim}}
\AddToHook{env/proposition/begin}{\crefalias{theorem}{proposition}}
\AddToHook{env/corollary/begin}{\crefalias{theorem}{corollary}}
\AddToHook{env/question/begin}{\crefalias{theorem}{question}}
\AddToHook{env/conjecture/begin}{\crefalias{theorem}{conjecture}}
\AddToHook{env/definition/begin}{\crefalias{theorem}{definition}}
\AddToHook{env/remark/begin}{\crefalias{theorem}{remark}}
\AddToHook{env/example/begin}{\crefalias{theorem}{example}}

\crefname{theorem}{Theorem}{Theorems}
\crefname{lemma}{Lemma}{Lemmas}
\crefname{proposition}{Proposition}{Propositions}
\crefname{claim}{Claim}{Claims}
\crefname{corollary}{Corollary}{Corollaries}
\crefname{question}{Question}{Questions}
\crefname{conjecture}{Conjecture}{Conjectures}
\crefname{definition}{Definition}{Definitions}
\crefname{remark}{Remark}{Remarks}
\crefname{example}{Example}{Examples}

\title{Upper Bounds for CSP Non-redundancy\\ via Locally Decodable Codes}
\author{Joshua Brakensiek\thanks{Department of Electrical Engineering and Computer Sciences, University of California, Berkeley 94720. Contact: \email{josh.brakensiek@berkeley.edu}}
\and David Conlon\thanks{Department of Mathematics, California Institute of Technology, Pasadena, CA 91125. Email: \email{dconlon@caltech.edu}}
\and Venkatesan Guruswami\thanks{Simons Institute for the Theory of Computing, and the Departments of EECS and Mathematics, University of California, Berkeley 94720. Contact: \email{venkatg@berkeley.edu}}
\and Xiaoyu He\thanks{School of Mathematics, Georgia Institute of Technology, Atlanta, GA 30332. Email: \email{xhe399@gatech.edu}}
\and Jiaxi Nie\thanks{School of Mathematics, Georgia Institute of Technology, Atlanta, GA 30332. Email: \email{jnie47@gatech.edu}}
\and Maya Sankar \thanks{Department of Mathematics, Massachusetts Institute of Technology, Cambridge, MA 02139. Email: \email{mayars@mit.edu}}
\and Hung-Hsun Hans Yu \thanks{Department of Mathematics, Princeton University, Princeton, NJ 08540. Email: \email{hansonyu@princeton.edu}}
}
\date{}
\begin{document}

\maketitle
\thispagestyle{empty}
\begin{abstract}
In the study of constraint satisfaction problems (CSPs), the non-redundancy of a relation measures the maximum number of clauses in a CSP instance which can be logically independent. In the last few years, CSP non-redundancy has found a number of applications including in query complexity, sparsification, kernelization and streaming.

In this paper, we present a novel technique for upper-bounding the non-redundancy of some relations using \emph{locally decodable codes} (LDCs) and \emph{locally correctable codes} (LCCs). More precisely, known limitations on the structure of 2-query LDCs and LCCs can be translated into limitations on the size of certain non-redundant CSP instances. This technique builds on a technique used by Brakensiek and Guruswami [STOC 2025] to bound non-redundancy via techniques from matching vector codes.

Using LDCs and LCCs, we give improved upper bounds on non-redundancy of the BCK predicate pioneered by Bessiere, Carbonnel, and Katsirelos [AAAI 2020] as well as its well-studied Boolean variant. We also give the first example of a Boolean predicate whose non-redundancy is provably not of the form $n^{k\pm o(1)}$ for some positive integer $k$.
\end{abstract}

\pagebreak

\section{Introduction}

An emerging field in the study of constraint satisfaction problems (CSPs) is systematic understanding of the \emph{non-redundancy} of various defining relations. First coined by Bessiere, Carbonnel, and Katsirelos~\cite{bessiere2020Chain} to capture the VC-dimension of a query complexity problem for CSPs, the notion has since found much success in characterizing other quantities related to CSPs such as their kernelization~\cite{carbonnel2022Redundancy,lagerkvist2020Sparsification,chen2020BestCase}, sparsification~\cite{khanna2024Code,khanna2024Characterizations,brakensiek2024Redundancy}, and streaming complexity~\cite{SV26a}. As such, there has been much motivation recently to study CSP non-redundancy as a standalone topic of interest~\cite{carbonnel2022Redundancy,khanna2024Characterizations,brakensiek2024Redundancy,brakensiek2025richness,SV26b}.

For a predicate $R \subseteq D^r$ of arity $r$ over (finite) domain $D$,  its non-redundancy on $n$ variables, denoted by $\NRD(R, n)$, is the maximal number of $R$-clauses one can impose on $n$ variables such that each clause is ``independently falsifiable.'' That is, for each clause in the instance, there is an assignment which satisfies every clause \emph{except} that one. As a canonical example, for the equality predicate $\EQ_D := \{(d,d) \mid d \in D\}$, to show that an instance (graph) is non-redundant, one must identify a cut which separates each clause (edge) a non-redundant. As such, non-redundant instances of $\EQ_D$ are precisely forests, so $\NRD(\EQ_D, n) = n-1$.

As another example, consider the predicate $\OR = \{(0,1),(1,0),(1,1)\}$. Here it turns out the complete (i.e., all pairs) instance is non-redundant as one can falsify any clause by setting its two variables to $0$ while satisfying all other clauses by setting every other variables to be $1$. Thus, $\NRD(\OR, n) = \binom{n}{2}$. The work of Bessiere, Carbonnel, and Katsirelos~\cite{bessiere2020Chain} (see also \cite{filtser2017Sparsification,butti2020Sparsificationa}) completely classified the non-redundant of all binary $(r=2)$ CSPs of any domain. In particular, all\footnote{We ignore the ``trivial'' predicates where $R = \emptyset$ or $R = D^r$.} such CSPs either have \emph{linear} non-redundancy--$\NRD(R, n) = \Theta(n)$--or \emph{quadratic} non-redundancy--$\NRD(R, n) = \Theta(n^2)$.

Despite there being such a simple classification for arity $2$ predicates, much less is known even for arity $r$. The $r=3$ case is fully classified in the Boolean setting, as always being linear, quadratic or cubic~\cite{chen2020BestCase,khanna2024Characterizations}, but for non-Boolean domains much more exotic behavior can occur. For example, Brakensiek and Guruswami~\cite{brakensiek2024Redundancy} identified a predicate $R \subseteq \{0,1,2\}^3$ for which $\NRD(R, n) \in [\Omega(n^{1.5}), O(n^{1.6}\log n)]$. That is, the dependence on $n$ in a non-redundancy function can be a non-integral polynomial. Furthermore,\cite{brakensiek2025richness} identifies other arity $3$ CSPs whose non-redundancy precisely corresponds to (unknown) tradeoffs between girth and edge-density in graphs (e.g., \cite{furedi2013history}). Other CSP predicates have been identified which closely relate to other landmark open problems in extremal combinatorics~\cite{carbonnel2022Redundancy,BrakensiekGP26}. As such, the study of CSP non-redundancy can be viewed as an important emergent test-bed for new techniques in logic and extremal combinatorics.

A key open question in the study of CSP non-redundancy is to understand the following predicate identified by Bessiere, Carbonnel, and Katsirelos~\cite{bessiere2020Chain} (first called the $\BCK$ predicate in \cite{brakensiek2024Redundancy}).
\[
  \BCK = \{(0,2,1),(1,0,2),(1,2,0),(2,0,1),(2,1,0)\} \subseteq \{0,1,2\}^3.
\]
That is, $\BCK$ consists of all permutations of $\{0,1,2\}$ except for the identify permutation. By relating the structure of $\BCK$ to the $(6,3)$ problem~\cite{ruzsa1978triple,fox2011new}, \cite{bessiere2020Chain} showed that $\NRD(\BCK, n) = O(n^2 / 2^{\log^* n})$. Such bounds were improved by \cite{brakensiek2024Redundancy} and \cite{brakensiek2025richness} to $O(n^{1.5}\log n)$ and $O(n^{1.5})$, respectively. Much effort has been put into trying to find a lower bound for $\NRD(\BCK, n)$ better than the trivial one of $\Omega(n)$, but no such attempt has succeeded (e.g., \cite{brakensiek2026superlinear}).

Among \emph{Boolean} non-redundancy problems, the frontier open question is a Boolean variant of $\BCK$, sometimes called $\BoolBCK$ in the literature, corresponding to all $3 \times 3$ permutation matrices except the identity matrix:\footnote{For succinctness, we denote long tuples as strings without commas or parentheses.}
\[
  \BoolBCK = \{100001010,010100001,010001100,001100010,001010100\},
\]
This predicate (or close variants) has appeared in a large number of works \cite{chen2020BestCase,lagerkvist2020Sparsification,khanna2024Characterizations,brakensiek2024Redundancy}, but the best known bounds on the non-redundancy of $\BoolBCK$ are
\[
  \NRD(\BoolBCK, n) = [\Omega(n), O(n^2)].
\]

\subsection{Our Contributions}\label{subsec:results}

Our main contribution in this paper is a novel technique for proving \emph{upper bounds} on CSP non-redundancy problems, which makes use of near-optimal bounds for 2-query locally decodable codes (LDCs) and locally correctable codes (LCCs). We describe the technique in more detail in Section~\ref{subsec:techniques}, but we first describe the new results we get with this technique.

As a first result, we give a new upper bound on the non-redundancy of $\BCK$, improving over the previous best bound of $O(n^{3/2})$~\cite{brakensiek2025richness}.

\begin{theorem}\label{thm:BCK}
$\NRD(\BCK, n) = O(n^{4/3})$.
\end{theorem}

As our second result, we give a new upper bound on the non-redundancy of $\BoolBCK$, the first to beat the $O(n^2)$ barrier~\cite{chen2020BestCase,lagerkvist2020Sparsification,khanna2024Characterizations,brakensiek2024Redundancy}.

\begin{theorem}\label{thm:BoolBCK}
$\NRD(\BoolBCK, n) = O(n^{3/2}\log^{5/2} n)$.
\end{theorem}

For both $\BCK$ and $\BoolBCK$, we cannot improve on the trivial lower bound of $\Omega(n)$. However, for some closely related predicates first considered by \cite{brakensiek2026superlinear}, we can apply our new upper-bound techniques to predicates with an existing lower bound. Namely, consider
\begin{align*}
  R_1 &= \{100001010,010100000,010001100,001100010,001010100,\\
        &\ \ \ \ \ \ 100010001,100001011,010100001,010001101,001100011,001010101\}\\
  R_2 &= \{100001000,010100010,010001100,001100000,001010100,\\
        &\ \ \ \ \ \ 100010011,100001001,010100011,010001101,001100001,001010101\}.
\end{align*}

Roughly speaking, these predicates are derived by deleting a coordinate (a.k.a. \emph{puncturing}) $\BoolBCK$ and then applying a product trick introduced in \cite{brakensiek2025richness}. %
For both of these predicates, by utilizing a result of \cite{brakensiek2026superlinear}, we can show that $\NRD(R_i, n) = \Omega(n^{11/5})$; see \cref{thm:R-bounds} for details.  Here, we complement these lower bounds with non-trivial upper bounds.

\begin{theorem}\label{thm:bool-fraction}
$\NRD(R_1, n) \le n^{8/3}\log^{O(1)} n$ and $\NRD(R_2,n)\le n^{13/5}\log^{O(1)}n.$
\end{theorem}

In particular, $R_1$ and $R_2$ are the \textbf{first} examples of \emph{Boolean} predicates whose non-redundancy has an asymptotically \emph{non-integral exponent}. This supplements existing results on non-Boolean domains by \cite{brakensiek2024Redundancy,brakensiek2025richness} as well as a result by Brakensiek, Guruswami, and Putterman~\cite{BrakensiekGP26} which gives an example of a Boolean predicate $R$ (of arity $4$) for which $\NRD(R, n) = n^{3-o(1)}$.

\subsection{Our Techniques}\label{subsec:techniques}

We now give a brief overview of the primary technical ideas behind our main results and how they compare to prior work. To begin, we first discuss how Brakensiek and Guruswami~\cite{brakensiek2024Redundancy} proved that $\NRD(\BCK, n) = O(n^{3/2}\log n)$. Consider the largest possible non-redundant instance of $\BCK = \{021,102,120,201,210\}$ on $n$ variables, which we number as $[n] := \{1,2, \hdots, n\}$. Concretely, this means there is a collection of edges $E \subseteq [n]^3$ such that for every hyperedge $e = (e_1, e_2, e_3)\in E$, there is a \emph{witness map} $\psi_e : [n] \to \{0,1,2\}$ such the the following criteria hold.

\begin{itemize}
\item[(1)] For every $e' = (e'_1, e'_2, e'_3) \in E \setminus \{e\}$, we have that $\psi_e(e') = (\psi_e(e'_1), \psi_e(e'_2), \psi_e(e'_3)) \in \BCK$.
\item[(2)] $\psi_e(e) = (\psi_e(e_1), \psi_e(e_2), \psi_e(e_3)) \not\in \BCK$.
\end{itemize}
In fact, by employing a trick known as \emph{conditional non-redundancy}, one can assume (essentially) without loss of generality that $\psi_e(e) = 012$, the identity permutation. We also assume without loss of generality (up to a constant factor) that $E$ is a \emph{tripartite hypergraph}. That is, there is a partition $V_1 \cup V_2 \cup V_3$ of $[n]$ such that $E \subseteq V_1 \times V_2 \times V_3$.

Given this family of solutions $\{\psi_e\}_{e \in E}$, consider an $[n] \times E$ matrix, where each column is $(\psi_e(1), \hdots, \psi_e(n))$ for some $e \in E$. We then label the rows of this matrix as $r_1, \hdots, r_n \in \{0,1,2\}^E$. A crucial observation by Brakensiek and Guruswami is that these rows could be interpreted as linear algebraic objects by embedding them in a finite vector space $\F_q^E$ via maps $\kappa_1, \hdots, \kappa_n : \{0,1,2\} \to \F_q$ such that the row $r_i = (r_{i,e})_{e \in E}$ becomes $v_i := (\kappa_i(r_{i,e}))_{e \in E} \in \F_q^E$. Importantly, this can be done in such a way such that for any hyperedge $e = (e_1, e_2, e_3) \in E$, we have that $v_{e_1} + v_{e_2} + v_{e_3} = 0$. However, by virtue of properties (1) and (2) above, for any hyperedge $e \in E$, the triple of vectors $(v_{e_1}, v_{e_2}, v_{e_3})$ looks sufficiently different from any other triples of vectors (coming from some hyperedge). In particular, for any $e \in E$ one can find a linear map $\sigma_e : \F_q^{E} \to \F_q$ such that $(\sigma_e(v_{e_1}), \sigma_e(v_{e_2}), \sigma_e(v_{e_3})) = (0, 0, 0)$ but $(\sigma_e(v_{e'_1}), \sigma_e(v_{e'_2}), \sigma_e(v_{e'_3})) \neq (0, 0, 0)$ for all $e' \in E \setminus \{e\}$.

Such a combination of linear and combinatorial constraints is highly reminiscent of \emph{matching vector codes}~\cite{yekhanin20083query,raghavendra2007note,gopalan2009note,efremenko20093query,dvir2011Matching}. By adapting techniques from the matching vector code literature, the authors of \cite{brakensiek2024Redundancy} managed to bound the size of $|E|$ by using the \emph{dimension} $d$ of the space spanned by the vectors $v_1, \hdots, v_n$. In particular, by exploiting a certain cubic polynomial representing $\BCK$, the authors get an upper bound of $|E| \le d^3$, which is sufficient to get the desired bound when $d = O(\sqrt{n})$. In the regime where $d \gg \sqrt{n}$, the authors prove using \emph{ad hoc} combinatorial methods a bound of (essentially) $|E| \le \frac{n^2 \log n}{d}$, which when balanced with $|E| \le d^3$ gets a desired bound of $O(n^{3/2}\log^{3/4} n)$.

\paragraph{The LDC/LCC Perspective.} The first conceptual contribution is that much of the work done by Brakensiek and Guruswami can be streamlined using the theory of \emph{locally decodable codes} (LDCs) and \emph{locally correctable codes} (LCCs). More precisely, let $M \in \F_q^{[n] \times E}$ be the matrix whose rows are $v_1, \hdots, v_n$, and let $C \subseteq \F_q^{[n]}$ be the linear space (a.k.a., linear code) spanned by the columns of $M$---that is, $M$ is the generator matrix of $C$. Now, the combinatorial properties of the vectors $v_1, \hdots, v_n$ observed earlier become much more transparently into local correctability of the code $C$. More precisely, the condition before that $v_{e_1} + v_{e_2} + v_{e_3} = 0$ now says that for any codeword $c \in C$, we have that $c_{e_1} + c_{e_2} + c_{e_3} = 0$. In particular, if the symbol $e_1$ were erased, it could be recovered from the values for $e_2$ and $e_3$. As such, $C$ is precisely a locally correctable code!

To complete the analysis of $\NRD(\BCK, n)$ from this perspective, we need to understand what fraction of errors $C$ can reliably decode from, which is based on the minimum degree of $E$ (but can be turned into average degree with a standard peeling argument). Using established bounds on the structure of LDCs (see Lemma~\ref{lem:LDC}), we can thus relate the number of edge $|E|$ to the dimension of the code $C$, allowing one to remove the \emph{ad hoc} arguments from  \cite{brakensiek2024Redundancy} while still recovering their bound.

The LDC/LCC perspective however gives us additional modularity which lets us improve the roughly $n^{3/2}$ bound of \cite{brakensiek2024Redundancy} to a significantly better bound of $n^{4/3}$ in \cref{thm:BCK}. This is done by relaxing the requirement that each variable of the original instance corresponds to a single coordinate of the corresponding code $C$. Instead, we give a more clever encoding where each variable of the original instance becomes \emph{three} coordinates in the corresponding code. Then, the cubic polynomial previously considered by \cite{brakensiek2024Redundancy} can be replaced by a quadratic polynomial. Furthermore, by embedding into the infinite field $\R$ rather than a finite field, sharp bounds on LCCs (see Lemma~\ref{lem:LCC}) can be used to obtain a crisp bound of $O(n^{4/3})$.

\paragraph{Boolean Instances.} Another way we distinguish our work from \cite{brakensiek2024Redundancy} is we are able to extend our technique to Boolean CSPs, particularly $\BoolBCK$. By analyzing a certain quadratic polynomial representing $\BoolBCK$, previous works have been able to show that $\NRD(\BoolBCK, n) = O(n^2)$~\cite{chen2020BestCase,khanna2024Characterizations,brakensiek2024Redundancy}, but none have been able to surpass that until our Theorem~\ref{thm:BoolBCK}. The primary reason is that $\BoolBCK$ is a predicate of arity $9$, making its behavior much more difficult to control.

Recall that $\BoolBCK$ corresponds to the five $3 \times 3$ non-identity permutation matrices. By doing a combinatorial transformation similar to the case of $\BCK$, it suffices to understand $9$-uniform hypergraphs $H = ([n], E)$ such that there exists a $9$-partition $[n] = \bigcup_{i=1}^3 \bigcup_{j=1}^3 V_{i,j}$ of the vertices such that $E \subseteq \prod_{i=1}^3 \prod_{j=1}^3 V_{i,j}$. Furthermore, for every hyperedge $e \in E$, there is a witness map $f_e : [n] \to \{0,1\}$ such that $f_e(e)$ is the identity matrix but for all $e' \in E \setminus \{e\}$, we have that $f_e(e')$ is a non-identity permutation matrix. By applying tricks similar to the proof of \cref{thm:BCK}, we get a code $C \subseteq \R^n$ corresponding to our witness maps. The key difference now is that an edge $e$ no longer corresponds to a single local constraint in the code $C$. Rather, if we view each edge $e$ as a $3 \times 3$ matrix, each row and column of $e$ (e.g., $(e_{1,1}, e_{1,2}, e_{1,3})$ is a column) corresponds to a local constraint in $C$. This makes counting the number of local constraints much more technical as many edges $e \in E$ can share the same row or column.

Analyzing the structure of this code uses a variety of algebraic and combinatorial techniques. On the algebraic side, by invoking \cref{lem:LDC}, we can bound $|E|$ in terms of the dimension of $C$ and the frequencies certain rows and columns appear. Furthermore, by using a certain quadratic representation of $\BoolBCK$, we can also bound that $|E| \le (\dim C)^2$. On the combinatorial side, if one restricts to studying the structure of induced $2 \times 2$ matrices, then some surprising tree-like structure emerges (see \cref{eq:9-unif-3/2-partial-first-bound}, allowing for the frequencies of the rows and columns to be related back to $|E|$. Combining these bounds yield \cref{thm:BoolBCK}.

The proof of \cref{thm:bool-fraction} is more technical than the proof of \cref{thm:BoolBCK} but the same primary techniques are used. As a preprocessing step, techniques in \cite{brakensiek2025richness,BrakensiekGP26} are used to transform $\NRD(R_1, n)$ and $\NRD(R_2, n)$ into questions about colorings of $8$-uniform hypergraphs. Roughly speaking, these hypergraphs closely correspond to the $9$-uniform hypergraphs considered in the proof of \cref{thm:BoolBCK}, except one of the $9$ partitions of the $9$-partite hypergraph is deleted---corresponding to either a ``diagonal'' or ``off-diagonal'' entry being erased in the corresponding permutation matrix. Once in this form, the techniques used to prove \cref{thm:BoolBCK} can largely be adapted, except for the fact that the missing partition greatly reduced the number of linear equations we have. We show that the remaining equations can largely ``pick up the slack,'' but there is some loss is incurred in making such adjustments. As a result, we get a bound with asymptotics closer to $n^{5/3}$ and $n^{8/5}$ (which translates to a bound of roughly $n^{8/3}$ and $n^{13/5}$ in \cref{thm:bool-fraction}) rather than $n^{3/2}$.

\paragraph{AI Usage Disclosure.} ChatGPT 5.5 Pro was used in the early stage of the project, although it did not lead to any progress.
It was used later to quickly summarize ideas of the authors, during which it discovered a simplification that led to the part titled ``Forbidding cycles'' in the proofs of \cref{lem:9-uniform-3/2} and \cref{lem:S1-5/3}.
All text in this paper was written by the authors.

\paragraph{Acknowledgments.}
The authors all thank the hospitality of the American Institute of Mathematics in Pasadena, CA, where this collaboration originated. This project emerged as an inter-SQuaRE collaboration between the ``Hypergraph Ramsey and Turán Theory'' SQuaRE and the ``Relaxations for promise CSPs'' SQuaRE.

We thank Zeev Dvir for suggesting the use of LCCs over a field of characteristic zero to shave a polylog factor in \cref{thm:BCK}. JB and VG thank Aaron Putterman for valuable discussions and encouragement.

JB and VG are supported in part by a Simons Investigator award and NSF grant CCF-2211972. JB is also supported in part by NSF grant DMS-2503280. DC is supported by NSF Grant DMS-2348859. HY is supported in part by a Jane Street Graduate Research Fellowship. MS was supported in part by NSF Grant DMS-2424441.

\section{Preliminaries}

\subsection{Notation, CSPs, and Non-redundancy}
We will use $A \lesssim B$ to denote $A=O(B)$. A hypergraph is linear if no two edges share more than one vertex. A hypergraph is $r$-uniform if every hyperedge has exactly $r$ vertices. Such an $r$-uniform hypergraph is further $r$-partite if there is a partition $V_1, \hdots, V_r$ of the vertex set $V$ such that every hyperedge contains exactly one vertex from each $V_i$ for all $i \in [r]$. 

We define a \emph{relation} (or \emph{predicate}) $R$ to be a subset of $D^r$, where $r \ge 1$ is the \emph{arity} of the predicate and $D$ is a finite \emph{domain}. We say that $R$ is \emph{non-trivial} if $R$ is neither empty nor $D^r$ itself. 

\begin{definition}
Given a predicate $R \subseteq D^r$, we define an \emph{instance} of $\CSP(R)$ to be a (directed) $r$-uniform hypergraph $(V, E)$ with $E \subseteq V^r$. We call each vertex $v \in V$ a \emph{variable} and each edge $e \in E$ a \emph{clause} or \emph{constraint}. We define an assignment to this instance to be a map $f : V \to D$. We say that a clause $(v_1, \hdots, v_r)$ is \emph{$R$-satisfied} (or just \emph{satisfied}) if $(f(v_1), \hdots, f(v_r)) \in R$. Otherwise, we say that the clause is \emph{$R$-unsatisfied}.
\end{definition}

\begin{definition}
Fix a non-trivial predicate $R \subseteq D^r$. Given an instance $(V, E)$ of $\CSP(R)$, we say that $(V, E)$ is \emph{non-redundant} if for every $e \in E$, there exists a \emph{witness} assignment $f_e : V \to D$ such that $f_e$ does not satisfy $e$ but satisfies every $e' \in E \setminus \{e\}$. We let $\NRD(R, n)$ denote the maximum number of edges in any non-redundant instance on $n$ variables. 
\end{definition}

We also define \emph{conditional non-redundancy} as coined by \cite{brakensiek2024Redundancy}, taking inspiration from \cite{bessiere2020Chain}. We define a \emph{promise relation} $P \mid Q$ to be a pair of predicates of the same arity and domain for which $P \subsetneq Q \subseteq D^r$.

\begin{definition}
We say that an $r$-uniform hypergraph $(V, E)$ is a non-redundant instance of $\CSP(P \mid Q)$ if for all $e \in E$, there exists a \emph{witness} assignment $f_e : V \to D$ satisfying the following properties: (1) every $e' \in E$ is $Q$-satisfied by $f_e$, (2) every $e' \in E \setminus \{e\}$ is $P$-satisfied by $f_e$, and (3) $e$ is \emph{not} $P$-satisfied by $f_e$. We let $\NRD(P \mid Q, n)$ denote the maximum size of a non-redundant instance on $n$ varibles subject to these constraints.
\end{definition}

Observe that $\NRD(P \mid D^r, n) = \NRD(P, n)$. We now list a few results in the literature on bounding the non-redundancy of (conditional) predicates. The first is a \emph{triangle inequality} due to \cite{brakensiek2024Redundancy}.

\begin{proposition}[Proposition 7.3 of \cite{brakensiek2024Redundancy}]\label{prop:triangle}
For any predicates $P \subsetneq Q \subsetneq R \subseteq D^r$, we have that
\[
    \NRD(P \mid Q, n) \le \NRD(P \mid R, n) \le \NRD(P \mid Q, n) + \NRD(Q \mid R, n).
\]
In particular, in the special case when $R = D^r$, we have that
\[
    \NRD(P \mid Q, n) \le \NRD(P, n) \le \NRD(P \mid Q, n) + \NRD(Q, n).
\]
\end{proposition}

Second, we describe how any conditional predicate can be mapped to a non-conditional predicate while changing the non-redundancy in a controlled manner. This technique was introduced in \cite{brakensiek2025richness} but here we use a more general formulation due to \cite{BrakensiekGP26}.

\begin{proposition}[Adapted from \cite{BrakensiekGP26}]\label{prop:decondition}
Assume $|D| \ge 2$. Consider any $P \subsetneq Q \subseteq D^r$. Pick distinct $a_0, a_1 \in D$ and let $R = P \times \{a_0\} \cup Q \times \{a_1\}$. Then
\[
    \Omega(n \cdot \NRD(P \mid Q, n)) \le \NRD(R, n) \le O(n \cdot \NRD(P \mid Q, n) + \NRD(Q \times \{a_0, a_1\}, n)).
\]
\end{proposition}
\begin{remark}
Technically, \cite{BrakensiekGP26} only prove the result for $r=3$, but the same proof goes through nearly verbatim for any arity $r$.
\end{remark}

\subsection{Locally Decodable and Locally Correctable Codes}
In this paper, the locally decodable codes and locally correctable codes we construct are going to take a particularly simple form.
We will thus restrict our attention to those codes.
For a more detailed introduction to general locally decodable and correctable codes, see \cite{yekhanin2012Locally,gopi2018Locality}.

\begin{definition}\label{def:LDC}
    A \emph{simple $2$-query linear LDC} over a field $\bF$ with error $\delta$, block length $n$ and message length $k$ consists of the following data.
    \begin{enumerate}
        \item Vectors $v_1,\ldots, v_n\in \bF^k$ with $v_1,\ldots, v_k$ spanning $\bF^k$.
        \item For each $i\in [k]$, a matching $M_i$ on $[n]\backslash \{i\}$ of size at least $\frac{\delta n}{2}$.
    \end{enumerate}
    Those data must satisfy the condition that $v_a-v_b$ is a nonzero multiple of $v_i$ for every $i\in [k]$ and $\{a,b\}\in M_i$.
\end{definition}
We remark that usually for $2$-query linear LDCs, we just require $v_i\in \Span\{v_a,v_b\}$.
However, any such code can be made simple with a small loss as shown by Dvir and Shpilka \cite{dvir2005locally}.

We also remark that given the data in \cref{def:LDC}, the actual LDC is defined by the encoding map $C:\bF^k\to \bF^n$ with $C(u)_i = u^{\intercal}Av_i$ for all $i\in [n]$.
Here, $A$ is the invertible $k$-by-$k$ matrix so that $Av_i$ is the $i$-th coordinate vector in $\bF^k$ for all $i\in [k]$.
Indeed, for any $i\in [k]$ and $\{a,b\}\in M_i$, suppose that $v_a-v_b = \alpha v_i$ for some $\alpha\in \bF^{\times}$.
Then for any $u\in\bF^k$, we have $u_i = C(u)_i = \alpha^{-1}(u^{\intercal}A)(v_a-v_b) = \alpha^{-1}C(u)_a-\alpha^{-1}C(u)_b$.
This shows that to recover the $i$-th bit of $u$, knowing the $a$-th position and the $b$-th position of $C(u)$ suffices.
Although the encoding-map perspective is closer to the usual notion of LDC, we choose to use an alternative equivalent definition for simple $2$-query linear LDCs as that is closer to how it is used in the proofs later.

Locally correctable codes are stronger locally decodable codes for which bits of codewords can also be recovered.
As for simple $2$-query linear LDCs, instead of the usual codeword perspective, we use the following equivalent formulation.
\begin{definition}\label{def:LCC}
    A \emph{simple $2$-query linear locally correctable code} over $\bF$ with error $\delta$, block length $n$ and message length $k$ consists of the following data.
    \begin{enumerate}
        \item Vectors $v_1,\ldots, v_n\in \bF^k$ spanning $\bF^k$.
        \item For each $i\in [n]$, a matching $M_i$ on $[n]\backslash \{i\}$ of size at least $\frac{\delta n}{2}$.
    \end{enumerate}
    Those data should satisfy the condition that $v_i\in\Span\{v_a,v_b\}$ for every $i\in [n]$ and $\{a,b\}\in M_i$.
    Additionally, we require that no edges appear in two different matchings.
\end{definition}
Given the data in \cref{def:LCC}, the actual locally correctable code $C\subseteq \bF^n$ consists of codewords spanned by $u_1,\ldots, u_k\in \bF^n$ where $(u_i)_j$ is the $i$-th coordinate of $v_j$ for every $i\in[k],j\in [n]$. That is, if we list the $v_i$'s as rows of a matrix, $C$ is spanned by its columns.
Indeed, if $i\in [n]$ and $\{a,b\}\in M_i$ satisfy $v_i = \alpha v_a+\beta v_b$, then any codeword $w\in C$ satisfies $w_i = \alpha w_a+\beta w_b$, showing that the $a$-th and $b$-th coordinates suffice to recover the $i$-th coordinate.

Having introduced the relevant notions of LDC and LCC for this paper, we now introduce the bounds we will need in the proofs.
For simple $2$-query linear LDCs, we can upper bound $k$ in terms of $n$ and $\delta$.

\begin{lemma}\label{lem:LDC}
    Any simple $2$-query linear LDC with error $\delta$, block length $n$ and message length $k$ satisfies
    \[k\leq \frac{\log_2 n}{\delta}.\]
\end{lemma}
We note that, up to a constant, the lemma was proven by Dvir and Shpilka in \cite{dvir2005locally} building on \cite{goldreich2006lower}, which proved the statement over $\bF_2$. We nonetheless provide a proof of \cref{lem:LDC} to be self-contained. 
\begin{proof}
    For each $s=0,1,\ldots, k$, let $G_s$ be the graph with vertex set $[n]$ and edge set $\bigcup_{i=1}^{s}M_i$.
    Let $V_1^{(s)},\ldots, V_{t_s}^{(s)}$ be the connected components of $G_s$.
    Note that from the definition of a simple $2$-query linear locally decodable code, it is clear that for any $s\in \{0,\ldots, k\}$, $\ell\in [t_s]$ and $i,j\in V_{\ell}^{(s)}$, we must have
    \[v_j-v_i\in \Span \{v_1,\ldots, v_s\}.\]

    Throughout the proof, we will keep track of the sizes of the connected components.
    To do so, define the potential function 
    \[\Phi(G_s) = \sum_{\ell=1}^{t_s}\card{V_{\ell}^{(s)}}\log_2\card{V_{\ell}^{(s)}}.\]
    We will show, for each $s\in [k]$, that
    \begin{equation}\label{eq:potential-induction}
        \Phi(G_s)\geq \Phi(G_{s-1})+2\card{M_s}.
    \end{equation}
    Assuming \cref{eq:potential-induction}, we will then have
    \[n\log_2 n \geq \Phi(G_k)\geq \Phi(G_0)+2\sum_{S=1}^{k}\card{M_s}\geq 0+\delta nk,\]
    which rearranges to the desired inequality.

    To compare $\Phi(G_s)$ and $\Phi(G_{s-1})$, we first restrict our attention on a single connected component $V_{\ell}^{(s)}$ of $G_s$.
    Suppose that it splits into $t$ connected components $V_{\ell_1}^{(s-1)},\ldots, V_{\ell_t}^{(s-1)}$ in $G_{s-1}$.
    We will show that
    \begin{equation}\label{eq:potential-induction-small-piece}
        \card{V_{\ell}^{(s)}}\log_2\card{V_{\ell}^{(s)}}\geq \sum_{i=1}^{t}\card{V_{\ell_i}^{(s-1)}}\log_2\card{V_{\ell_i}^{(s-1)}} + \card{M_s|_{V_{\ell}^{(s)}}}
    \end{equation}
    where $M_s|_{V_{\ell}^{(s)}}$ consists of edges of $M_s$ within $V_{\ell}^{(s)}$.
    Summing this over all connected components of $G_s$ then gives \cref{eq:potential-induction}.
    
    If $t=1$, we claim that $M_s$ has no edges within $V_{\ell}^{(s)}$.
    Suppose for the contradiction that $a,b\in V_{\ell}^{(s)}$ span an edge in $M_s$.
    Then $v_a-v_b$ is a nonzero multiple of $v_s$ by assumption.
    But we also know that $v_a-v_b\in \Span\{v_1,\ldots, v_{s-1}\}$ as $a$ and $b$ are connected in $G_{s-1}$.
    This shows that $v_s\in \Span\{v_1,\ldots, v_{s-1}\}$, contradicting the linear independence of the vectors $v_1,\ldots, v_k$. 
    Therefore $M_s|_{V_{\ell}^{(s)}}$ is empty, proving \cref{eq:potential-induction-small-piece} when $t=1$.

    For $t\geq 2$, by the concavity of $\log_2$ on $(0,\infty)$ and Jensen's inequality, we have
    \[\sum_{i=1}^{t}\card{V_{\ell_i}^{(s-1)}}\log_2\card{V_{\ell_i}^{(s-1)}}\leq \card{V_{\ell}^{(s)}}\log_2\frac{\card{V_{\ell}^{(s)}}}{t}\leq \card{V_{\ell}^{(s)}}\log_2\card{V_{\ell}^{(s)}}-\card{V_{\ell}^{(s)}}.\]
    On the other hand, $M_s|_{V_{\ell}^{(s)}}$ is a matching in $V_{\ell}^{(s)}$, showing that
    \[2\card{M_s|_{V_{\ell}^{(s)}}}\leq \card{V_{\ell}^{(s)}}.\]
    Summing the two inequalities up finish the proof of \cref{eq:potential-induction-small-piece} for $t\geq 2$, and hence the proof of the lemma.
\end{proof}

Since simple $2$-query linear LCCs are, after some small modifications, also simple $2$-query linear locally decodable codes, the bound in \cref{lem:LDC} also holds for them with some extra loss.
The bound in \cref{lem:LDC} can indeed be tight over $\bF_2$, for example when $\delta=1$ and $v_1,\ldots, v_n$ are all the vectors in $\bF_2^k$.
However, over $\bR$, we can save an extra logarithmic factor for simple $2$-query linear locally correctable codes using a result of Dvir, Saraf and Wigderson.
For the reader's convenience, we specialize and state their result here.

\begin{lemma}[{\cite[Corollary 1.4]{dvir2014improved}}]\label{lem:DSW14}
Let $m,n,k$ be positive integers.
Let $A$ be an $m$-by-$n$ real matrix satisfying the following properties:
\begin{enumerate}
    \item each row of $A$ has at most $3$ nonzero entries;
    \item each column of $A$ has at least $k$ nonzero entries;
    \item for any two columns, there are at most $3$ rows at which the two columns are simultaneously nonzero.
\end{enumerate}
Then the rank of $A$ is at least $n-18n/k$.
\end{lemma}

\begin{lemma}\label{lem:LCC}
    Any simple $2$-query linear LCC over $\bR$ with error $\delta$, block length $n$ and message length $k$ satisfies $k\le 36\delta^{-1}$.
\end{lemma}
\begin{proof}
    We will build a matrix $A$ from any given simple $2$-query linear locally correctable code and apply \cref{lem:DSW14}.
    The columns of $A$ are indexed by $[n]$ and the rows are indexed by the pair $(i,\{a,b\})$ where $i\in [n]$ and $\{a,b\}\in M_i$.

    For any row index $(i,\{a,b\})$, we know that $v_i = \alpha v_a+\beta v_b$ for some $\alpha,\beta\in \bR^{\times}$.
    The matrix $A$ is defined so that the row vector at $(i,\{a,b\})$ is $\alpha e_a+\beta e_b-e_i$, where $e_j$ is the $j$-th coordinate vector for any $j\in[n]$.
    This is defined so that if $B$ is the $n$-by-$k$ matrix whose row vectors are $v_1,\ldots, v_n$, then $AB=0$.

    Now it is clear that each row of $A$ has precisely three nonzero entries.
    Moreover, the $i$-th column has at least $\card{M_i}\geq \frac{\delta n}{2}$ nonzero entries.
    Lastly, for any $i\neq i'\in [n]$, suppose that the $i$-th column and the $i'$-th column are simultaneously nonzero at row $(j,\{a,b\})$.
    Then $i,i'\in \{j,a,b\}$.

    If $j=i$, then $i'\in \{a,b\}$, which occurs at most once as $M_i$ is a matching.
    Similarly $j=i'$ can only occur at most once.
    Lastly, if $j\neq i,i'$, then we must have $\{a,b\}=\{i,i'\}$.
    As we require each edge to appear in at most once matching, this also occurs at most once.
    Therefore the two columns can only be simultaneously nonzero at at most three rows.

    \cref{lem:DSW14} now shows that the rank of $A$ is at least $n-36\delta^{-1}$.
    As an immediate consequence, the (right) kernel of $A$ has dimension at most $36\delta^{-1}$.
    As $AB=0$, this shows that the column space of $B$ has dimension at most $36\delta^{-1}$, and thus $\dim \Span\{v_1,\ldots, v_n\}\leq 36\delta^{-1}$.
    By assumption, $v_1,\ldots, v_n$ span the entire $\bR^k$, showing that $k\leq 36\delta^{-1}$, as desired.
\end{proof}

\begin{remark}
    In the definition of simple $2$-query LCCs, we require each edge to be present in at most one matching.
    Without that assumption, the best known bound in the literature is currently $k = O(\delta^{-9})$ by \cite{barak2011Rank}.
    Even using the later improvement in \cite{dvir2014improved}, the argument would only give $k=O(\delta^{-6})$.
    In our applications, both bounds are inferior to \cref{lem:LDC} as $\delta$ is polynomial in $n$.
\end{remark}

\section{Proof of \cref{thm:BCK}}\label{sec:BCK}

Our first step is to convert our non-redundancy problem for $\BCK$ into an extremal combinatorics problem which can be analyzed using locally correctable codes. This first step is a mild adaptation of the techniques in \cite{brakensiek2024Redundancy}.

Define $\BCK^{+} := \{(0,1,2), (0,2,1), (1,0,2), (1,2,0), (2,0,1), (2,1,0)\}$ be the set of all permutations of $\{0, 1, 2\}$. Note that $\BCK \subseteq \BCK^{+}$. Furthermore, it is known that $\NRD(\BCK^{+}, n) = O(n)$, see \cite{brakensiek2024Redundancy}. Thus, by applying \cref{prop:triangle}, we have 
\[
    \NRD(\BCK, n) \le \NRD(\BCK \mid \BCK^{+}, n) + \NRD(\BCK^{+}, n) = \NRD(\BCK \mid \BCK^{+}, n) + O(n).
\]
In other words, to prove \cref{thm:BCK}, it suffices to show that $\NRD(\BCK \mid \BCK^{+}, n) = O(n^{4/3})$.

To simplify the problem further, we make use of a multipartite trick (e.g., \cite{brakensiek2025richness}). Let $(V, E)$ be a non-redundant instance of $\BCK \mid \BCK^{+}$. Define the \emph{$3$-lift} of $(V, E)$ to be the $3$-partite hypergraph $(V', E')$ where $V' = V \times \{3\}$ and
\[
  E' = \{((e_1, 1), (e_2, 2), (e_3, 3)) \mid (e_1, e_2, e_3) \in E\}.
\]
By design $(V', E')$ is $3$-partite, has the same number of edges as $(V, E)$ but $|V'| = 3|V|$. One can show that if $(V, E)$ is non-redundant, then $(V', E')$ is also non-redundant.\footnote{In more detail, any assignment $\psi : V \to D$ for $(V, E)$ corresponds to the ``forgetful'' assignment $\psi' : V' \to D$ where $\psi'(v, i) = \psi(v)$. Thus, the witnesses that $(V, E)$ is non-redundant map to witnesses that $(V', E')$ is non-redundant.} Thus, to prove $\NRD(\BCK \mid \BCK^{+}, n) = O(n^{4/3})$ (and thus \cref{thm:BCK}), it suffices to prove the following statement in extremal combinatorics.

\begin{lemma}\label{lem:3-uniform-4/3}
Let $H=(V, E)$ be a $3$-uniform hypergraph such that there exists a tripartition $V = V_1 \cup V_2 \cup V_3$ of the vertices such that $E \subseteq V_1 \times V_2 \times V_3$. Further assume for each $e \in E$, there is a map $f_e : V \to \{0,1,2\}$ such that $f_e(e) = (0, 1, 2)$ but for all $e' \in E \setminus \{e\}$, we have that $f_e(e') \in \BCK = \{(0,2,1), (1,0,2), (1,2,0), (2,0,1), (2,1,0)\}$. Then, $|E| = O(|V|^{4/3})$. 
\end{lemma}

\begin{proof}
First of all, we note that $H$ must be linear.
Suppose for the sake of contradiction that $\card{e\cap e'}=2$ for some two edges $e,e'$.
Without loss of generality, suppose that $e_1=e_1'$ and $e_2=e_2'$.
As $f_e(e) = (0,1,2)$, we know that $f_e(e_1)=0$ and $f_e(e_2)=1$.
However, no matter what $f_e(e_3')$ is, it is impossible for $f_e(e')$ to lie in $\BCK$ as the projection of $\BCK$ to the first two coordinates avoids $(0,1)$.
Therefore $H$ must be linear.

We first trim the hypergraph $H$ so that its minimum degree is large.
To do so, let $d = \card{E}/2\card{V}$.
Whenever there is a vertex with degree less than $d$, remove that vertex and all edges incident to it.
Let $H' = (V',E')$ be the resulting hypergraph.
Then
\[\card{E'}\geq \card{E}-d\card{V}\geq \frac{1}{2}\card{E}.\]
Moreover, we know that $\deg_{H'}v \geq d$ for every $v\in V'$.

Now for any $v\in V'$, define three vectors $x_{(v,0)}, x_{(v,1)}, x_{(v,2)}\in \{0,1\}^{E'}\subseteq \bR^{E'}$ as follows: for any $i\in \{0,1,2\}$ and $e\in {E'}$, set $(x_{(v,i)})_e = 1_{f_e(v)=i}.$
Then for every $e,e'\in E'$ and every $i\in \{0,1,2\}$, we clearly have
\[(x_{(e_1,i)})_{e'}+(x_{(e_2,i)})_{e'}+(x_{(e_3,i)})_{e'}=1\]
as every element of $\BCK^+$ has precisely one coordinate equal to $i$.
Thus $x_{(e_1,i)}+x_{(e_2,i)}+x_{(e_3,i)} = \vec{1}$, where $\vec{1}\in \bR^{E'}$ is the all-one vector.

To make those vectors a simple 2-query LCC, let $\pi:\bR^{E'}\to \bR^{\card{E'}-1}$ be any surjective linear map whose kernel is the span of $\vec{1}$.
Also fix a choice $i\in \{0,1,2\}$, and set $W_i = \Span\{x_{v,i}:v\in V'\}$.
Then $\pi(x_{(e_1,i)})+\pi(x_{(e_2,i)})+\pi(x_{(e_3,i)})=0$ for any $e \in E'$.
For any $v\in V'$, let $M_v = \{\{u,w\}:u,v,w\textup{ form an edge in }E'\}.$
Since $H$ is linear, $H'$ is linear as well, showing that $(M_v)_{v\in V'}$ are pairwise edge-disjoint matchings.
Hence, the vectors $(\pi(x_{(v,i)}))_{v\in V'}$ together with the matchings $(M_v)_{v\in V'}$ form a simple $2$-query LCC.
Its error parameter can be bounded below by $d/\card{V'}\geq d/\card{V}$, and its message length is $\dim \pi(W_i)$.
By \cref{lem:LCC}, we thus have 
\begin{equation}\label{eq:3-uniform-dim-bound}
    \dim W_i\leq \dim \pi(W_i)+1\lesssim \frac{\card{V}}{d}\lesssim  \frac{\card{V}^2}{\card{E}}.
\end{equation}

For any $w_0\in W_0$ and $w_1\in W_1$, define $\phi(w_0,w_1)$ so that $\phi(w_0,w_1)_e = (w_0)_e\cdot (w_1)_e$ for all $e\in E'$.
In other words, $\phi(w_0,w_1)$ is the Hadamard product of the two vectors $w_0$ and $w_1$.
Since $\phi:W_0\times W_1\to \bR^{E'}$ is bilinear, we know that $\dim \phi(W_0\times W_1)\leq \dim (W_0\otimes W_1) = \dim W_0\dim W_1$.

Now for every $e\in E'$, set $y_e = \phi(x_{(e_1,0)},x_{(e_2,1)})\in \phi(W_0\times W_1)$.
For every $e'\in E'$, we know that $\phi(x_{(v_1,0)},x_{(v_2,1)})_{e'}=0$ or $1$, and it is $1$ if and only if both $x_{(v_1,0)},x_{(v_2,1)}$ are both $1$ at $e'$.
The latter is equivalent to $f_{e'}(e_1)=0$, $f_{e'}(e_2)=1$, which is equivalent to $e_1=e_1'$ and $e_2=e_2'$ by the assumption.
Since $H$ is linear, this is equivalent to $e=e'$.
Therefore $y_e$ is the unit vector in the $e$-th direction, and so $\phi(W_0\times W_1)$ contains all the unit vectors in $\bR^{E'}$.
As a consequence,
\[\frac{1}{2}\card{E}\leq \card{E'} = \dim\phi(W_0\times W_1)\leq \dim W_0\dim W_1\lesssim \frac{\card{V}^4}{\card{E}^2}\]
where we use \cref{eq:3-uniform-dim-bound} for the last inequality.
Rearranging gives that $|E| =O(|V|^{4/3})$.
\end{proof}

\begin{remark}
The use of LCCs rather than LDCs in this proof is purely to sharpen the inequality. Using only the LDC bound of \cref{lem:LDC}, we can prove $|E| \le |V|^{4/3}\log^{O(1)}|V|$, which is still a polynomial improvement over the state of the art.
\end{remark}

\section{Proof of \cref{thm:BoolBCK}}

To prove \cref{thm:BoolBCK}, we begin by converting our CSP non-redundancy problem into a combinatorics problem analogous to the proof of \cref{thm:BCK}. Let
\[
    \BoolBCKp := \{100010001, 100001010,010100001,010001100,001100010,001010100\}
\]
be the set of all $3 \times 3$ permutation matrices, written as an elements of $\{0,1\}^9$. By results of \cite{lagerkvist2020Sparsification,chen2020BestCase,khanna2024Characterizations,brakensiek2024Redundancy}, we know that $\NRD(\BoolBCKp, n) = O(n)$. Since $\BoolBCK \subseteq \BoolBCKp$, by \cref{prop:triangle}, we have that
\begin{align*}
    \NRD(\BoolBCK, n) &\le \NRD(\BoolBCK \mid \BoolBCK^{+}, n) + \NRD(\BoolBCK^{+}, n)\\
    &= \NRD(\BoolBCK \mid \BoolBCK^{+}, n) + O(n).
\end{align*}
Thus, it suffices to prove that $\NRD(\BoolBCK \mid \BoolBCK^{+}, n) = O(n^{3/2})$. Similar to the argument in \cref{sec:BCK}, with only a constant-factor loss in the instance size, we may assume our non-redundant instance of $\CSP(\BoolBCK \mid \BoolBCK^{+})$ is $9$-partite. As it is more convenient to index these parts as entries in a $3 \times 3$ matrix, we obtain the following extremal combinatorics problem.

\begin{lemma}\label{lem:9-uniform-3/2}
Let $H=(V, E)$ be a $9$-uniform hypergraph such that there exists a $9$-partition $V = \bigcup_{i=1}^3 \bigcup_{j=1}^3 V_{i,j}$ of the vertices such that $E \subseteq \prod_{i=1}^3 \prod_{j=1}^3 V_{i,j}$. Further assume for each $e \in E$, there is a map $f_e : V \to \{0,1\}$ such that $f_e(e)$ is the identity matrix but for all $e' \in E \setminus \{e\}$, we have that $f_e(e')$ is a non-identity permutation matrix. Then, $\card{E}=O( |V|^{3/2}\log^{5/2} |V|)$.  
\end{lemma}

\begin{proof}
Since the proof is significantly more involved, we divide the proof into several pieces.

\paragraph{Making the shadow linear.}
As in the proof of \cref{lem:3-uniform-4/3}, we first clean up the $9$-uniform hypergraph.
This time, the cleaning process is more complicated and will alter the hypergraph $H$ itself by contracting vertices.

Define $S$ to be the $3$-uniform hypergraph on $V$ as follows: for each edge $e$ in $H$, put in the edges $(e_{i,1},e_{i,2},e_{i,3})$ and $(e_{1,i},e_{2,i},e_{3,i})$ for each $i\in [3]$ in $S$.
In other words, it is hypergraph recording the ``shadows'' of the edges in $H$ in each row and column.
We will contract vertices in $H$ so that the corresponding shadow hypergraph $S$ is linear.

Suppose that two edges in $S$ share exactly two vertices.
Without loss of generality, assume that they are $(e_{1,1},e_{1,2},e_{1,3})$ and $(e_{1,1}',e_{1,2}',e_{1,3}')$ with $e_{1,1}=e_{1,1}'$ and $e_{1,2}=e_{1,2}'$.
As any element in $\BoolBCKp$, when put in matrix form, has rows and columns summing to one, we have
\[f_{e''}(e_{1,3}) = 1-f_{e''}(e_{1,1})-f_{e''}(e_{1,2}) = f_{e''}(e_{1,3}')\]
for any $e''\in E$.
Therefore we can contract $e_{1,3}$ and $e_{1,3}'$ while keeping $(f_{e''})_{e''\in E}$ well-defined.
As such contraction decreases the number of vertices by $1$, eventually $S$ must become linear.
Moreover, the number of edges remains unchanged throughout the process, so the task is still to prove the same upper bound on $\card{E}$.

\paragraph{Dyadic decomposition.} Set $\ell = \lfloor \log_2 \card{V}\rfloor+1$.
For any $t\in [\ell]$ and $i\in[3]$, let $V_{i,i}^{(t)}$ be the set of vertices $v\in V_{i,i}$ with $\deg_Sv \in [2^{t-1}, 2^t)$.
For any $t_1,t_2,t_3\in [\ell]$, let $E^{(t_1,t_2,t_3)}$ be the set of edges $e\in E$ with $e_{1,1}\in V_{1,1}^{(t_1)}$, $e_{2,2}\in V_{2,2}^{(t_2)}$ and $e_{3,3}\in V_{3,3}^{(t_3)}$.
It is clear that
\[E = \bigcup_{t_1,t_2,t_2\in [\ell]}E^{(t_1,t_2,t_3)}\]
by definition, and we will bound each $\card{E^{(t_1,t_2,t_3)}}$ separately.
Therefore, we will fix the choice of $t_1,t_2,t_3\in [\ell]$ until the last step.
We will bound $\card{E^{(t_1,t_2,t_3)}}$ in two different ways---one using an argument similar to the proof of \cref{lem:3-uniform-4/3}, and another elementary one via identifying forbidden structures. The former bound is most effective when the degree buckets $t_1, t_2, t_3$ are large, while the latter bound is most effective in the small-degree case. To get the best overall bounds, we carefully combine both inequalities using Cauchy-Schwarz.

\paragraph{Forbidding cycles.} We start by showing that cycles are forbidden in some graph obtained by projecting to the $e_{1,1}$- and $e_{2,2}$-coordinates.
For any $a\in V_{1,2}$ and $b\in V_{2,1}$, let $E^{(t_1,t_2,t_3)}_{a,b}$ be the set of edges $e\in E^{(t_1,t_2,t_3)}$ with $e_{1,2}=a$ and $e_{2,1}=b$.
Let $G_{a,b}$ be a bipartite graph on $V^{(t_1)}_{1,1}\cup V^{(t_2)}_{2,2}$ whose edges are projection of $E^{(t_1,t_2,t_3)}_{a,b}$ to the corresponding coordinates.
We first observe that this projection is an injection.
Indeed, if $e,e'$ are sent to the same edges, then $e_{i,j} = e_{i,j}'$ for any $i,j\in[2]$.
By the linearity of $S$, we have $e_{3,1}=e_{3,1}'$, $e_{3,2}=e_{3,2}'$,  $e_{1,3}= e_{1,3}'$ and $e_{2,3}=e_{2,3}'$.
This forces $e_{3,3}=e_{3,3}'$ and so $e=e'$ again by the linearity of $S$.
Moreover, if a vertex has positive degree in $G_{a,b}$, then it must be in some edge with both $a$ and $b$.
Set $U^{(t_1)}_{a,b}$ to be the vertices in $V^{(t_1)}_{1,1}$ with positive degrees, and $W^{(t_2)}_{a,b}$ to be the vertices in $V^{(t_2)}_{2,2}$ with positive degrees as well.

The key claim is that $G_{a,b}$ is a forest.
Suppose that it contains a cycle $u_1w_1u_2w_2\cdots u_sw_s$ where $u_1,\ldots, u_s\in U^{(t_1)}_{a,b}$ and $w_1,\ldots, w_s\in W^{(t_2)}_{a,b}$.
Let $e$ be the edge containing $u_1,a,b,w_1$.
Then by the assumption, we know that $f_e(u_1)=f_e(w_1)=1$.
We also know that $f_e(a)=f_e(b)=0$.
Now for any other edges $e'\neq e$ containing $u\in V_{1,1}^{(t_1)},a,b,w\in V_{2,2}^{(t_2)}$, it is easy to check that $(f_{e}(u),f_{e}(w))$ must either be $(0,1)$ or $(1,0)$.
Therefore we may prove by induction that $f_e(u_2)=\cdots =f_e(u_s)=0$ and $f_e(w_2)=\cdots=f_e(w_s)=1$.
However, $(f_e(u_1),f_e(w_s))=(1,1)$, which is a contradiction.
Therefore
\[\card{E^{(t_1,t_2,t_3)}_{a,b}} = \card{E(G_{a,b})}< \card{U^{(t_1)}_{a,b}}+\card{W^{(t_2)}_{a,b}}.\]
As a consequence,
\begin{equation}\label{eq:9-unif-3/2-partial-first-bound}
    \card{E^{(t_1,t_2,t_3)}}<\sum_{a\in V_{1,2}, b\in V_{2,1}}\left(\card{U^{(t_1)}_{a,b}}+\card{W^{(t_2)}_{a,b}}\right).
\end{equation}
Now for each $u\in V_{1,1}^{(t_1)}$, the number of choices of $(a,b)$ for $U^{(t_1)}_{a,b}$ to contain $u$ can be upper bounded by $(\deg_S u)^2 < 2^{2t_1}$.
Therefore
\[\sum_{a\in V_{1,2}, b\in V_{2,1}}\card{U^{(t_1)}_{a,b}}\leq  2^{2t_1}\card{V^{(t_1)}_{1,1}},\]
and similarly
\[\sum_{a\in V_{1,2}, b\in V_{2,1}}\card{W^{(t_2)}_{a,b}}\leq  2^{2t_2}\card{V^{(t_2)}_{2,2}}.\]
With \cref{eq:9-unif-3/2-partial-first-bound}, this shows that
\begin{equation}\label{eq:9-unif-3/2-first-bound}
    \card{E^{(t_1,t_2,t_3)}}\leq 2^{2t_1}\card{V_{1,1}^{(t_1)}}+2^{2t_2}\card{V_{2,2}^{(t_2)}}.
\end{equation}

\paragraph{Building a LDC.} Now let us bound $\card{E^{(t_1,t_2,t_3)}}$ using ideas from \cref{lem:3-uniform-4/3}.
Similar to the proof of \cref{lem:3-uniform-4/3}, we will associate each vertex with a vector recording the values of $(f_e)_{e\in E}$.
This time, we can work over arbitrary fields.
For simplicity, let us work over $\bF_2$.
For each $v\in V$, let $x_v\in \bF_2^{E^{(t_1,t_2,t_3)}}$ be the vector with $(x_v)_e = f_e(v)$ for every $e\in E^{(t_1,t_2,t_3)}$.
We also write $\pi$ as the linear map from $\bF_2^{E^{(t_1,t_2,t_3)}}$ with $\Span\{\vec{1}\}$ being its kernel.

Now suppose that $(v_{2,1},v_{2,2},v_{2,3})$ is an edge in $S$.
Then there exists an edge $e\in E$ with $e_{2,i}=v_{2,i}$ for every $i\in[3]$.
For any $e'\in E$, we know that $f_{e'}(e)$ must be a permutation matrix, showing that
\[f_{e'}(v_{2,1})+f_{e'}(v_{2,2})+f_{e'}(v_{2,3})=1.\]
Therefore
\[x_{v_{2,1}}+x_{v_{2,2}}+x_{v_{2,3}} = \vec{1}.\]
This shows that $\pi(x_{v_{2,2}}) = \pi(x_{v_{2,1}})-\pi(x_{v_{2,3}})$ as we are working over $\bF_2$.
Similarly, if $(v_{1,2},v_{2,2},v_{3,2})$ is an edge in $S$, then $\pi(x_{v_{2,2}}) = \pi(x_{v_{1,2}})-\pi(x_{v_{3,2}})$.
Therefore we are very close to getting a simple $2$-query LDC over $\bF_2$.

To actually produce a simple $2$-query LDC, let 
\[X_2^{(t_2)} = \Span\{x_{v_{2,2}}:v_{2,2}\in V_{2,2}^{(t_2)}\}\]
and let $u_1,\ldots, u_k\in V_{2,2}^{(t_2)}$ be such that $(\pi(x_{u_i}))_{i\in [k]}$ form a basis of $\pi(X_2^{(t_2)})$.
Let $\pi'$ be any projection map from the image of $\pi$ to $\pi(X_2^{(t_2)})$, and write $\sigma = \pi'\circ \pi$.
Finally, for any $i\in [k]$, let $M_i$ be the pairs of vertices that form an edge with $u_i$ in $S$.
It is then clear that $(\sigma(x_v))_{v\in V}$ together with the matchings $(M_i)_{i\in [k]}$ give a simple $2$-query LDC with block length $\card{V}$ and message length $k$.
Moreover, since 
\[\card{M_i} = \deg_S u_i\geq 2^{t_2-1}\]
for every $i\in [k]$, we can set the error parameter as $2^{t_2}/\card{V}$.
\cref{lem:LDC} then shows that
\[\dim \pi(X_2^{(t_2)}) = k \leq \frac{\card{V}\log_2\card{V}}{2^{t_2}}.\]
Therefore
\begin{equation}\label{eq:9-unif-3/2-dim-bound-X2}
    \dim X_2^{(t_2)} = \dim \pi(X_2^{(t_2)})+1\lesssim \frac{\card{V}\log\card{V}}{2^{t_2}}.
\end{equation}
If we set $X_3^{(t_3)} = \Span \{x_{v_{3,3}}:v_{3,3}\in V_{3,3}^{(t_3)}\}$, we also have
\begin{equation}\label{eq:9-unif-3/2-dim-bound-X3}
  \dim X_3^{(t_3)} \lesssim \frac{\card{V}\log\card{V}}{2^{t_3}}  
\end{equation}
using the same argument.

As in the proof of \cref{lem:3-uniform-4/3}, we now let $\phi:X_2^{(t_2)}\times X_3^{(t_3)}\to \bF_2^{E^{(t_1,t_2,t_3)}}$ be such that $\phi(x_2,x_3)_e = (x_2)_e\cdot (x_3)_e$ for every $e\in E^{(t_1,t_2,t_3)}$.
For any edge $e\in E^{(t_1,t_2,t_3)}$, we know that $e_{2,2}\in V_{2,2}^{(t_2)}$ and $e_{3,3}\in V_{3,3}^{(t_3)}$ by definition, which shows that $(x_{e_{2,2}},x_{e_{3,3}})\in X_2^{(t_2)}\times X_3^{(t_3)}$.
Moreover, it is clear that $\phi(x_{e_{2,2}},x_{e_{3,3}})$ is the coordinate vector corresponding to $e$.
This shows that the image of $\phi$ spans the entire space, which shows that $\phi$ is surjective as $\phi$ is bilinear.
By considering the dimensions of the vector spaces, we obtain that
\begin{equation}\label{eq:9-unif-3/2-second-bound}
    \card{E^{(t_1,t_2,t_3)}}\leq \dim X_2^{(t_2)}\dim X_3^{(t_3)}\lesssim \frac{\card{V}^2\log^2\card{V}}{2^{t_2+t_3}}
\end{equation}
using \cref{eq:9-unif-3/2-dim-bound-X2} and \cref{eq:9-unif-3/2-dim-bound-X3}.

\paragraph{Final steps.} Multiplying \cref{eq:9-unif-3/2-first-bound} and \cref{eq:9-unif-3/2-second-bound} now gives
\[\card{E^{(t_1,t_2,t_3)}}^2\lesssim \left(2^{t_1-t_3}\card{V_{1,1}^{(t_1)}}+2^{t_2-t_3}\card{V_{2,2}^{(t_2)}}\right)\card{V}^2\log^2\card{V},\]
and so
\begin{equation}\label{eq:9-unif-3/2-individual-bound}
    \card{E^{(t_1,t_2,t_3)}}\lesssim \left(\sqrt{2^{t_1-t_3}\card{V_{1,1}^{(t_1)}}}+\sqrt{2^{t_2-t_3}\card{V_{2,2}^{(t_2)}}}\right)\card{V}\log\card{V}.
\end{equation}

We now sum this bound over all $t_1,t_2,t_3\in [\ell]$ with $t_1\leq t_2\leq t_3$.
The other cases can be similarly bounded, which would only enlarge the upper bound by a factor of six.
We may now compute
\[\sum_{1\leq t_1\leq t_2\leq t_3\leq \ell}\sqrt{2^{t_2-t_3}\card{V_{2,2}^{(t_2)}}} \leq \ell \sum_{1\leq t_2\leq \ell}\sum_{t_2\leq t_3\leq \ell}\sqrt{2^{t_2-t_3}\card{V_{2,2}^{(t_2)}}}\lesssim \ell \sum_{1\leq t_2\leq \ell}\sqrt{\card{V_{2,2}^{(t_2)}}}< \ell \sqrt{\ell \card{V}}\]
where the last inequality follows from an application of Cauchy--Schwarz.
Similarly we also have
\[\sum_{1\leq t_1\leq t_2\leq t_3} \sqrt{2^{t_1-t_3}\card{V_{1,1}^{(t_1)}}} \lesssim \ell\sqrt{\ell \card{V}}.\]
Plugging the last two inequalities into \cref{eq:9-unif-3/2-individual-bound}, we have
\[\card{E}\lesssim \ell\sqrt{\ell\card{V}} \cdot \card{V}\log \card{V} \lesssim \card{V}^{3/2}\log^{5/2}\card{V}.\qedhere\]
\end{proof}

\section{Proof of \cref{thm:bool-fraction}}

\subsection{Punctured Permutation Matrices}

The most important step toward proving \cref{thm:bool-fraction} is to prove variants of \cref{lem:9-uniform-3/2} for ``punctured'' permutation matrices. To be more precise, given a subset $S \subseteq \{1,2,3\} \times \{1,2,3\}$, we say that a map $f : S \to \{0,1\}$ is a punctured permutation matrix if $f$ is a restriction of a permutation matrix map $g : \{1,2,3\}^2 \to \{0,1\}$. Likewise, $f$ is a punctured identity matrix if $f$ is a restriction of $g(i,j) = \mathbf{1}[i = j].$ Among puncturings that remove a single element from $\{1,2,3\} \times \{1,2,3\}$, without loss of generality there are two posibilites, the \emph{diagonal} pucturing $S_1 := \{1,2,3\}^2 \setminus \{(3,3)\}$ and the \emph{off-diagonal} puncturing $S_2 := \{1,2,3\}^2 \setminus \{(2,3)\}$. In either case, we can set up a natural variant of \cref{lem:9-uniform-3/2}.

\begin{lemma}\label{lem:S1-5/3}
Let $(V, E)$ be a $8$-uniform hypergraph such that there exists a $8$-partition $V = \bigcup_{(i,j) \in S_1} V_{i,j}$ of the vertices such that $E \subseteq \prod_{(i,j) \in S_1} V_{i,j}$. Further assume for each $e \in E$, there is a map $f_e : V \to \{0,1\}$ such that $f_e(e)$ is a punctured identity matrix but for all $e' \in E \setminus \{e\}$, we have that $f_e(e')$ is a non-identity punctured permutation matrix. Then, $\card{E}=O\left( |V|^{5/3}\log^{2} |V|\right)$. 
\end{lemma}

\begin{lemma}\label{lem:S2-8/5}
Let $(V, E)$ be a $8$-uniform hypergraph such that there exists a $8$-partition $V = \bigcup_{(i,j) \in S_2} V_{i,j}$ of the vertices such that $E \subseteq \prod_{(i,j) \in S_2} V_{i,j}$. Further assume for each $e \in E$, there is a map $f_e : V \to \{0,1\}$ such that $f_e(e)$ is a punctured identity matrix but for all $e' \in E \setminus \{e\}$, we have that $f_e(e')$ is a non-identity punctured permutation matrix. Then, $\card{E}=O\left( |V|^{8/5}\log^{21/5} |V|\right)$.  
\end{lemma}

The proofs of two lemmas are mostly analogous to the proof of \cref{lem:9-uniform-3/2}.
We start with the proof of \cref{lem:S1-5/3}, whose proof is the same as the proof of \cref{lem:9-uniform-3/2} except for some details.

\begin{proof}[Proof of {\cref{lem:S1-5/3}}]
We still divide the proof into five parts.

\paragraph{Making the shadow linear.} Since we do not have the full matrix anymore, we will define the shadow $3$-uniform hypergraph $S$ slightly differently.
For each edge $e$ in $H$, put in the edges $(e_{i,1},e_{i,2},e_{i,3})$ and $(e_{1,i},e_{2,i},e_{3,i})$ now only for $i=1,2$ as $e_{3,3}$ is absent.
The exact same argument as in \cref{lem:9-uniform-3/2} still allows us to assume $S$ is linear after finitely many contractions of vertices.

\paragraph{Dyadic decomposition.} We still set $\ell = \lfloor \log_2\card{V}\rfloor+1$.
For any $t\in [\ell]$ and $i\in [2]$, let $V^{(t)}_{i,i}$ be the set of vertices $v\in V_{i,i}$ with $\deg_Sv\in [2^{t-1},2^t)$.
For any $t_1,t_2\in [\ell]$, let $E^{(t_1,t_2)}$ be the set of edges $e\in E$ with $e_{1,1}\in V_{1,1}^{(t_1)}$ and $e_{2,2}\in V_{2,2}^{(t_2)}$.
We still have
\[E = \bigcup_{t_1,t_2\in [\ell]}E^{(t_1,t_2)},\]
and we will still fix $t_1,t_2\in [\ell]$ until the last step.
Before that, we will derive two different bounds on $\card{E^{(t_1,t_2)}}$.

\paragraph{Forbidding cycles.}
This part of the proof of \cref{lem:9-uniform-3/2} does not use the index $e_{3,3}$, and the proof works almost verbatim: we only need to change $E^{(t_1,t_2,t_3)}$ to $E^{(t_1,t_2)}$ and to omit $e_{3,3}=e_{3,3}'$ when checking the projection is injective.
Therefore we still have
\begin{equation}\label{eq:S1-5/3-first-bound}
    \card{E^{(t_1,t_2)}}\leq 2^{2t_1}\card{V_{1,1}^{(t_1)}}+2^{2t_2}\card{V_{2,2}^{(t_2)}}.
\end{equation}

\paragraph{Building LDC. } This part of the proof of \cref{lem:9-uniform-3/2} does use $e_{3,3}$, but it does not use $e_{1,1}$. After swapping $1$ and $3$ in all the indices and replacing $E^{(t_1,t_2,t_3)}$ with $E^{(t_1,t_2)}$, the proof in \cref{lem:9-uniform-3/2} works verbatim.
Thus we still have
\begin{equation}\label{eq:S1-5/3-second-bound}
    \card{E^{(t_1,t_2)}}\lesssim \frac{\card{V}^2\log^2\card{V}}{2^{t_1+t_2}}.
\end{equation}

\paragraph{Final steps. }This part is the only different part, although the main idea stays the same.
Multiplying \cref{eq:S1-5/3-first-bound} with the square of \cref{eq:S1-5/3-second-bound}, we get
\[\card{E^{(t_1,t_2)}}^3\lesssim \left(2^{-2t_2}\card{V_{1,1}^{(t_1)}}+2^{-2t_1}\card{V_{2,2}^{(t_2)}}\right)\card{V}^4\log^4\card{V},\]
and so
\begin{equation}\label{eq:S1-5/3-individual-bound}
    \card{E^{(t_1,t_2)}}\lesssim \left(2^{-2t_2/3}\card{V_{1,1}^{(t_1)}}^{1/3}+2^{-2t_1/3}\card{V_{2,2}^{(t_2)}}^{1/3}\right)\card{V}^{4/3}\log^{4/3}\card{V}.
\end{equation}
Now we know that
\[\sum_{t_1,t_2\in[\ell]}\left(2^{-2t_2/3}\card{V_{1,1}^{(t_1)}}^{1/3}+2^{-2t_1/3}\card{V_{2,2}^{(t_2)}}^{1/3}\right)\lesssim \sum_{t_1\in[\ell]}\card{V_{1,1}^{(t_1)}}^{1/3}+\sum_{t_2\in[\ell]}\card{V_{2,2}^{(t_2)}}^{1/3}\lesssim \ell^{2/3}\card{V}^{1/3}\]
by an application of H\"older's inequality.
Plugging this into \cref{eq:S1-5/3-individual-bound}, we obtain
\[\card{E} = \sum_{t_1,t_2\in[\ell]}\card{E^{(t_1,t_2)}}\lesssim \card{V}^{5/3}\log^2\card{V}.\qedhere\]

\end{proof}

The proof of \cref{lem:S2-8/5} is almost the same.
The main difference is that the forbidding cycles step no longer works, and this step will be replaced with a simple counting argument using the linearity of the shadow hypergraph.
To make the counting argument work, the dyadic decomposition will be modified as well.

\begin{proof}[Proof of \cref{lem:S2-8/5}]
The proof is still divided into five parts.
The third part is going to be a counting argument instead. %

\paragraph{Making the shadow linear.} 
As before, the shadow $3$-uniform hypergraph $S$ is slightly different as now $(2,3)$ is absent from the index set $S_2$.
For each edge in $e\in H$, put in the edges $(e_{i,1},e_{i,2},e_{i,3})$ and $(e_{1,j},e_{2,j},e_{3,j})$ for every $i\neq 2$ and $j\neq 3$.
The exact same argument as in \cref{lem:9-uniform-3/2} still shows that we may assume $S$ to be linear after finitely many contractions.

\paragraph{Dyadic decomposition.}
Still set $\ell = \lfloor\log_2\card{V}\rfloor+1$.
The dyadic decomposition will be done with respect to the index set $T = \{(3,2),(3,1),(1,2),(1,1)\}$.
For any $t\in [\ell]$ and $(i,j)\in T$, set $V_{i,j}^{(t)}$ to be the set of vertices $v\in V_{i,j}$ with $\deg_Sv\in [2^{t-1},2^t)$.
For any $t_1,t_2,t_3,t_4\in [\ell]$, let $E^{(t_1,t_2,t_3,t_4)}$ be the set of edges $e\in E$ with $e_{i,j}\in V_{i,j}^{(t_{k})}$ where $k=6-i-j$ for every $(i,j)\in T$. 
We still have 
\[E = \bigcup_{t_1,t_2,t_3,t_4\in [\ell]}E^{(t_1,t_2,t_3,t_4)}.\]
For the next two parts, we fix some $t_1,t_2,t_3,t_4\in [\ell]$ and bound the size of $E^{(t_1,t_2,t_3,t_4)}$ in two different ways.

\paragraph{Simple counting.}
Note that by the linearity of $S$, an edge $e\in E^{(t_1,t_2,t_3,t_4)}$ is uniquely determined by $(e_{i,j})_{(i,j)\in S_2\backslash {\{(1,3)\}}}$.
Now for each choice of $e_{3,2}\in V_{3,2}^{(t_1)}$, there are at most $2^{t_1}$ choices of $(e_{1,2},e_{2,2})$ and $2^{t_1}$ choices of $(e_{3,1},e_{3,3})$ by how the dyadic decomposition was done.
Once those five vertices are determined, there are at most $2^{t_2}$ choices for $(e_{1,1},e_{2,1})$ as $e_{3,1}\in V_{3,1}^{(t_2)}$.
Therefore 
\[\card{E^{(t_1,t_2,t_3,t_4)}}\leq 2^{2t_1+t_2}\card{V_{3,2}^{(t_1)}}.\]
Instead of counting $(e_{i,j})_{(i,j)\in S_2\backslash \{(1,3)\}}$, we can also count $(e_{i,j})_{(i,j)\in S_2\backslash \{(2,1)\}}$.
When the five vertices $(e_{3,2},e_{1,2},e_{2,2},e_{3,1},e_{3,3})$ are determined, there are at most $2^{t_3}$ choices for $(e_{1,1},e_{1,3})$ as $e_{1,2}\in V_{1,2}^{(t_3)}$.
Therefore we also have
\[\card{E^{(t_1,t_2,t_3,t_4)}}\leq 2^{2t_1+t_3}\card{V_{3,2}^{(t_1)}}.\]
As a consequence,
\[\card{E^{(t_1,t_2,t_3,t_4)}}\leq 2^{2t_1+\min\{t_2,t_3\}}\card{V_{3,2}^{(t_1)}}\leq 2^{2t_1+\min\{t_2,t_3\}}\card{V}.\]
Now, instead of starting from $e_{3,2}$, we may also start from $e_{1,1}$, extend it to $(e_{1,1},e_{1,2},e_{1,3},e_{2,1},e_{3,1})$ in at most $2^{2t_4}$ ways and then either choose $(e_{2,2},e_{3,2})$ or $(e_{3,2},e_{3,3})$ to determine $e$.
Therefore the same counting argument gives
\[\card{E^{(t_1,t_2,t_3,t_4)}}\leq 2^{2t_4+\min\{t_2,t_3\}}\card{V}.\]
We may combine the two bounds above as
\begin{equation}\label{eq:S2-8/5-counting}
    \card{E^{(t_1,t_2,t_3,t_4)}}\leq 2^{2\min\{t_1,t_4\}+\min\{t_2,t_3\}}\card{V}.
\end{equation}

\paragraph{Building LDC.} We still define for each $v\in V$ a vector $x_v\in \F_2^{E^{(t_1,t_2,t_3,t_4)}}$ with $(x_v)_e = f_e(v)$ for every $e\in E^{(t_1,t_2,t_3,t_4)}$.
Also for every $(i,j)\in T$ and $k=6-i-j$, let
\[X_{i,j}^{(t_k)} = \Span\left\{x_v:v\in V^{(t_k)}_{i,j}\right\}.\]
The argument in \cref{lem:9-uniform-3/2} still shows that
\begin{equation}\label{eq:S2-8/5-LDC}
    \dim X_{i,j}^{(t_k)}\lesssim \frac{\card{V}\log\card{V}}{2^{t_k}}.
\end{equation}
We will use \cref{eq:S2-8/5-LDC} to derive two different upper bounds on $\card{E^{(t_1,t_2,t_3,t_4)}}$.

Set $\phi:\F_2^{E^{(t_1,t_2,t_3,t_4)}}\times \F_2^{E^{(t_1,t_2,t_3,t_4)}}\to \F_2^{E^{(t_1,t_2,t_3,t_4)}}$ to be the map with $\phi(x,y)_e = x_e\cdot y_e$ for all $e\in E^{(t_1,t_2,t_3,t_4)}$.
We have seen that $\phi(x_{e_{2,2}},x_{e_{3,3}})$ is the coordinate vector corresponding to $e$ for each $e\in E^{(t_1,t_2,t_3,t_4)}$.
We will use $x_{e_{2,2}} = \vec{1}-x_{e_{3,2}}-x_{e_{1,2}}$ and $x_{e_{3,3}} = \vec{1}-x_{e_{3,2}}-x_{e_{3,1}}$ to expand $\phi(x_{e_{2,2}},x_{e_{3,3}})$.
Note that $\phi(\vec{1},x) = \phi(x,\vec{1}) = \phi(x,x)= x$ for every $x\in \F_2^{E^{(t_1,t_2,t_3,t_4)}}$.
Using this and the fact that we are working over $\F_2$, we get
\[\phi(x_{e_{2,2}},x_{e_{3,3}}) = \vec{1}+x_{e_{3,2}}+x_{e_{3,1}}+x_{e_{1,2}}+\phi(x_{e_{3,2}},x_{e_{3,1}})+\phi(x_{e_{1,2}},x_{e_{3,2}})+\phi(x_{e_{1,2}},x_{e_{3,1}}).\]
The right hand side is contained in 
\[\Span\left\{\{\vec{1}\}\cup X_{3,2}^{(t_1)}\cup X_{3,1}^{(t_2)}\cup X_{1,2}^{(t_3)}\cup \phi( X_{3,2}^{(t_1)}, X_{3,1}^{(t_2)})\cup \phi( X_{1,2}^{(t_3)},X_{3,2}^{(t_1)})\cup \phi(X_{1,2}^{(t_3)},X_{3,1}^{(t_2)})\right\},\]
whereas the left hand side runs through all coordinate vectors in $\F_2^{E^{(t_1,t_2,t_3,t_4)}}$ when $e$ runs through the edges in $E^{(t_1,t_2,t_3,t_4)}$.
Therefore
\begin{align}\label{eq:S2-8/5-first-LDC-bound}
    \card{E^{(t_1,t_2,t_3,t_4)}}\leq &1+\dim X_{3,2}^{(t_1)}+\dim X_{3,1}^{(t_2)}+\dim  X_{1,2}^{(t_3)}\nonumber\\&+\dim X_{3,2}^{(t_1)}\dim X_{3,1}^{(t_2)}+\dim X_{1,2}^{(t_3)}\dim X_{3,2}^{(t_1)}+\dim X_{1,2}^{(t_3)}\dim X_{3,1}^{(t_2)}\nonumber\\
    \lesssim & \frac{\card{V}^2\log^2\card{V}}{2^{\min\{t_1+t_2,t_1+t_3,t_2+t_3\}}}\leq \frac{\card{V}^2\log^2\card{V}}{2^{\min\{t_2,t_3\}+\min\{t_1,t_2,t_3,t_4\}}}
\end{align}
by \cref{eq:S2-8/5-LDC}.

For the other bound, we claim that $\phi(x_{e_{1,1}},\vec{1}-x_{e_{3,2}})$ is the coordinate vector corresponding to $e$ for each $e\in E^{(t_1,t_2,t_3,t_4)}$.
Indeed, for any $e'\in E^{(t_1,t_2,t_3,t_4)}$, we know that $(x_{e_{1,1}})_{e'}\cdot \left(1-x_{e_{3,2}}\right)_{e'}$ is either zero or one.
Moreover, it is one if and only if $f_{e'}(e_{1,1}) = 1$ and $f_{e'}(e_{3,2})=0$.
We clearly have $f_e(e_{1,1})=1$ and $f_e(e_{3,2})=0$.
On the other hand, if $f_{e'}(e_{1,1}) = 1$ and $f_{e'}(e_{3,2})=0$, then we must have $f_{e'}(e_{1,j})=f_{e'}(e_{i,1})=0$ for $i,j\in \{2,3\}$.
Since $f_{e'}(e_{1,2})= f_{e'}(e_{3,2})=0$, we have $f_{e'}(e_{2,2})=1$.
Similarly $f_{e'}(e_{3,1})= f_{e'}(e_{3,2})=0$ implies $f_{e'}(e_{3,3})=1$.
This shows that $f_{e'}(e)$ is a punctured identity matrix, and so $e'=e$.
As a result, $\phi(x_{e_{1,1}},\vec{1}-x_{e_{3,2}})$ is the coordinate vector corresponding to $e$.

Since $\phi(x_{e_{1,1}},\vec{1}-x_{e_{3,2}}) = x_{e_{1,1}}-\phi(x_{e_{1,1}},x_{e_{3,2}})$ is contained in the span of $X_{1,1}^{(t_4)}$ and $\phi(X_{1,1}^{(t_4)},X_{3,2}^{(t_1)})$, we have
\begin{equation}\label{eq:S2-8/5-second-LDC-bound}\card{E^{(t_1,t_2,t_3,t_4)}}\leq \dim X_{1,1}^{(t_4)}+\dim X_{1,1}^{(t_4)}\dim X_{3,2}^{(t_1)}\lesssim \frac{\card{V}^2\log^2\card{V}}{2^{t_1+t_4}}\leq \frac{\card{V}^2\log^2\card{V}}{2^{2\min\{t_1,t_4\}}}
\end{equation}
again by \cref{eq:S2-8/5-LDC}.

Note that $\max\{a+\min\{a,b\},2b\} = b+\max\{a,b\}$ for any real numbers $a,b$.
Therefore, by taking the better of the two bounds \cref{eq:S2-8/5-first-LDC-bound} and \cref{eq:S2-8/5-second-LDC-bound}, we get
\begin{equation}\label{eq:S2-8/5-combined-LDC-bound}
    \card{E^{(t_1,t_2,t_3,t_4)}}\lesssim \frac{\card{V}^2\log^2\card{V}}{2^{\min\{t_1,t_4\}}+2^{\max\{\min\{t_1,t_4\},\min\{t_2,t_3\}\}}}.
\end{equation}

\paragraph{Final steps.}
By multiplying the square of \cref{eq:S2-8/5-counting} with the third power of \cref{eq:S2-8/5-combined-LDC-bound}, we get
\[\card{E^{(t_1,t_2,t_3,t_4)}}^5\lesssim 2^{\min\{t_1,t_4\}+2\min\{t_2,t_3\}-3\max\{\min\{t_1,t_4\},\min\{t_2,t_3\}\}}\card{V}^8\log^6\card{V}.\]
To simplify, set $\alpha(t_1,t_2,t_3,t_4) = \min\{t_1,t_2,t_3,t_4\}$ and $\beta(t_1,t_2,t_3,t_4) = \max\{\min\{t_1,t_4\},\min\{t_2,t_3\}\}$.
Then we can rewrite the inequality as 
\[\card{E^{(t_1,t_2,t_3,t_4)}}\lesssim 2^{(\alpha(t_1,t_2,t_3,t_4)-\beta(t_1,t_2,t_3,t_4))/5}\card{V}^{8/5}\log^{6/5}\card{V}.\]
To sum this over $t_1,t_2,t_3,t_4\in [\ell]$, note that for any fixed $\alpha,\beta\in[\ell]$ with $\alpha \leq \beta$, there are at most $O(\ell^2)$ choices for $t_1,t_2,t_3,t_4$ so that $\alpha(t_1,t_2,t_3,t_4)=\alpha$ and $\beta(t_1,t_2,t_3,t_4)=\beta$.
Therefore
\[\card{E}\lesssim \sum_{1\leq \alpha\leq \beta\leq \ell}\ell^22^{(\alpha-\beta)/5}\card{V}^{8/5}\log^{6/5}\card{V}\lesssim \sum_{1\leq \alpha\leq \ell}\ell^2\card{V}^{8/5}\log^{6/5}\card{V}\lesssim \card{V}^{8/5}\log^{21/5}\card{V}.\qedhere\]
\end{proof}

\subsection{Completing the Proof of \cref{thm:bool-fraction}}
\label{}
To prove \cref{thm:bool-fraction}, we need to combine \cref{lem:S1-5/3} and \cref{lem:S2-8/5} with tools from CSP non-redundancy. To start with, we define the following predicates.
\begin{align*}
P_1 &= \{10000101,01010000,01000110,00110001,00101010\}\\
Q_1 &= \{10001000,10000101,01010000,01000110,00110001,00101010\}\\
P_2 &= \{10000100,01010001,01000110,00110000,00101010\}\\
Q_2 &= \{10001001,10000100,01010001,01000110,00110000,00101010\}
\end{align*}

By inspection, one can verify that $P_1 \subseteq Q_1$ and $P_2 \subseteq Q_2$. Furthermore, $R_1 = P_1 \times \{0\} \cup Q_1 \times \{1\}$ and $R_2 = P_2 \times \{0\} \cup Q_2 \times \{1\}$. By \cref{lem:S1-5/3} and \cref{lem:S2-8/5}, we can immediately deduce the following corollary.
\begin{corollary}\label{cor:PQi}
We have that $\NRD(P_1 \mid Q_1, n) = O(n^{5/3} \log^{2} n)$ and also $\NRD(P_2 \mid Q_2, n) = O(n^{8/5} \log^{21/5} n)$. 
\end{corollary}

Note that $P_1$ and $Q_1$ are precisely $\BoolBCK$ and $\BoolBCKp$, respectively, with the last coordinate removed. Furthermore, $P_2$ and $Q_2$ are precisely $\BoolBCK$ and $\BoolBCKp$, respectively, with the penultimate coordinate removed. As such, by \cite{brakensiek2026superlinear} we have the following lower bounds
\begin{proposition}[Theorem 22, \cite{brakensiek2026superlinear}]\label{prop:PQlb}
For $i \in [2]$, $\NRD(P_i \mid Q_i, n) = \Omega(n^{6/5})$.
\end{proposition}

The goal of this section is to combine \cref{cor:PQi} and \cref{prop:PQlb} to prove the following result concerning $R_1$ and $R_2$.
\begin{theorem}\label{thm:R-bounds}
We have that $\NRD(R_1, n) \in [\Omega(n^{11/5}), O(n^{8/3}\log^{2} n)]$ and that $\NRD(R_2, n) \in [\Omega(n^{11/5}), O(n^{13/5}\log^{21/5} n)]$. 
\end{theorem}
\begin{proof}
Fix $i \in [2]$. Since $R_i = P_i \times \{0\} \cup Q_i \times \{1\}$, we have by \cref{prop:decondition} that
\[
    \Omega(n \cdot \NRD(P_i \mid Q_i, n)) \le \NRD(R_i, n) \le O(n \cdot \NRD(P_i \mid Q_i, n) + \NRD(Q_i \times \{0, 1\}, n)).
\]
By plugging in \cref{cor:PQi} and \cref{prop:PQlb}, we thus get that
\[
    n^{11/5} \lesssim \NRD(R_1, n) \lesssim n^{8/3}\log^{2} n + \NRD(Q_1 \times \{0,1\}, n).
\]
and
\[
    n^{11/5} \lesssim \NRD(R_2, n) \lesssim n^{13/5}\log^{21/5} n + \NRD(Q_1 \times \{0,1\}, n).
\]

Thus, it suffices to prove that $\NRD(Q_i \times \{0,1\}, n) = O(n^2)$. A sufficient way to prove this (due to \cite{jansen2019Optimal,chen2020BestCase,khanna2024Characterizations}) is to identify two quadratic polynomials $q_1, q_2 : \R^9 \to \R$ such that for an evaluation point $x \in \{0,1\}^9$ we have that $q_i(x) = 0$ if and only if $x \in Q_i \times \{0,1\}$. A suitable choice of polynomials is as follows\footnote{We note that such polynomials can be found computationally using the lattice-based methods of \cite{khanna2024Characterizations,BrakensiekGP26}. We give \emph{ad hoc} arguments in this paper for completeness.}
\begin{align*}
    q_1(x_1, x_2, \hdots, x_9) &= (x_1+x_2+x_3-1)^2 + (x_4+x_5+x_6-1)^2 + (x_1+x_4+x_7-1)^2\\
    &\ \ \ + (x_2+x_5+x_8-1)^2 + (x_3+x_6-x_7-x_8)^2 + (x_3+x_6)(x_3+x_6-1)\\
    &\ \ \ + (x_7+x_8)(x_7+x_8-1),\\
    q_2(x_1, x_2, \hdots, x_9) &= (x_1+x_2+x_3-1)^2 + (x_4+x_5+x_6-1)^2 + (x_1+x_4+x_7-1)^2\\
    &\ \ \ + (x_3+x_6+x_8-1)^2 + (x_2+x_5-x_7-x_8)^2 + (x_3 + x_6)(x_3+x_6-1)\\
    &\ \ \ + (x_7+x_8)(x_7+x_8-1).
\end{align*}
Note that both polynomials ignore the input $x_9$. To analyze $q_1$, fix $x \in \{0,1\}^9$ and arrange the first $8$ coordiantes into the following matrix
\[
\begin{pmatrix}
x_1 & x_2 & x_3\\
x_4 & x_5 & x_6\\
x_7 & x_8 & -
\end{pmatrix}
\]
If $q_1(x) = 0$, then the first two rows and the first two columns must have exactly one $1$. Furthermore, the terms $(x_3+x_6)(x_3+x_6-1)$ and $(x_7+x_8)(x_7+x_8-1)$ enforce that the last row and column each have at most one $1$. Finally, $(x_3+x_6-x_7-x_8)^2$ imposes the requireement that the sum of the entries in the last row and the sum of the entries in the last column are equal. If we let $x'_9 = 1 - x_3 - x_6$, then $q_1(x) = 0$ if and only if
\[
\begin{pmatrix}
x_1 & x_2 & x_3\\
x_4 & x_5 & x_6\\
x_7 & x_8 & x'_9
\end{pmatrix}
\]
is a permutation matrix, which is equivalent to $(x_1, \hdots, x_8, x'_9) \in \BoolBCK$ which in turn is also equivalent to $(x_1, \hdots, x_8) \in Q_1$. Thus, $q_1(x) = 0$ if and only if $x \in Q_1 \times \{0,1\}$.

The argument for $q_2$ is analogous. Fix $x \in \{0,1\}^9$ and arrange the first $8$ coordinate into the following matrix
\[
\begin{pmatrix}
x_1 & x_2 & x_3\\
x_4 & x_5 & x_6\\
x_7 & - & x_8
\end{pmatrix}
\]
The terms of $q_2$ impose that the first two rows and the first and last columns of the above matrix have exactly one $1$. Furthermore, the second column and the last row each have at most one $1$ and this second column and last row have the same sum. Thus, if we let $x'_9 = 1 - x_2 - x_5$, we have that $q_2(x) = 0$ iff
\[
    \begin{pmatrix}
x_1 & x_2 & x_3\\
x_4 & x_5 & x_6\\
x_7 & x'_9 & x_8
\end{pmatrix}
\]
is a permutation matrix. In other words, $q_2(x) = 0$ iff $(x_1, \hdots, x_7, x'_9, x_8) \in \BoolBCK$ which is true iff $(x_1, \hdots, x_8) \in Q_2$. Therefore, $q_2(x) = 0$ if and only if $x \in Q_2 \times \{0,1\}$, as desired.
\end{proof}

Observe that \cref{thm:R-bounds} immediately implies \cref{thm:bool-fraction}.

\section{Conclusion}

In this paper, we show how techniques from the theory of locally decodable and locally correctable codes can be used to bound the non-redundancy of various constraint satisfaction problems.  Given the rich connections that have been found between CSP non-redundancy and various other computational problems involving CSPs~\cite{lagerkvist2020Sparsification,chen2020BestCase,bessiere2020Chain,carbonnel2022Redundancy,brakensiek2024Redundancy,SV26a}, our results also make advances on these other questions.

We conclude with two open questions. First, is \cref{thm:BCK} tight? If not, is it possible that $\NRD(\BCK, n) = O(n)$? If the latter possibility holds, then this would lead to a new class of CSPs which have near-linear-sized CSP sparsifiers \cite{khanna2024Code,khanna2024Characterizations,brakensiek2024Redundancy}. More precisely, the most general known argument for showing that a CSP has linear-sized non-redundancy is by invoking a generalized form a Gaussian elimination due to Bulatov and Dalmau~\cite{bulatov2006Simple,bessiere2020Chain,chen2020BestCase,lagerkvist2020Sparsification}. However, it is known that such methods cannot work for predicates like $\BCK$ due to barriers in universal algebra~\cite{bessiere2020Chain,brakensiek2025richness}. Getting a linear-sized bound for $\BCK$ via a different method would likely open up an entirely new perspective on CSP non-redundancy and sparsification. %

Second, this paper identifies a few specific problems for which the LDC/LCC method can be used to make progress. Can a more general theory be built to understand for which CSP predicates these tools are useful? Identifying the limit of these techniques could be a crucial step toward finding other methods of bounding the non-redundancy of CSPs.

\bibliographystyle{alpha}
\bibliography{main}

\end{document}